# Influence of tetragonal distortion on the topological electronic structure of the half-Heusler compound LaPtBi from first principles

X. M. Zhang,[1,3] W. H. Wang,[1, a)] E. K. Liu,[1] G. D. Liu,[3] Z. Y. Liu,[2] and G. H. Wu[1]

[1]*Beijing National Laboratory for Condensed Matter Physics, Institute of Physics, Chinese Academy of Sciences, Beijing 100080, P. R. China*

[2] *State Key Laboratory of Metastable Material Sciences and Technology, Yanshan University Technology, Qinhuangdao 066004, P. R. China*

[3] *School of Material Sciences and Engineering, Hebei University Technology, Tianjin 300130, P. R. China*

Abstract:

The electronic structures of tetragonally distorted half-Heuselr compound LaPtBi in the $C1_b$ structure are investigated in the framework of density functional theory using the full potential linearized augmented plane with local spin density approximation method. The calculation results show that both the band structures and the Fermi level can be tuned by using either compressive or tensile in-plane strain. A large bulk band gap of 0.3 eV can be induced through the application of a compressive in-pane strain in LaPtBi with the assumption of a relaxed volume of the unit cell. Our results could serve as a guidance to realize topological insulators in half-Heusler compounds by strain engineering.

---

a) Electronic mail: wenhong.wang@iphy.ac.cn

Topological insulator (TI) is of a new class of materials, which has a bulk band gap generated by strong spin-orbit coupling but contains gapless surface states. [1-3] These surface states are chiral and inherently robust to external perturbations because they are protected by time-reversal symmetry. Based on such unique electronic surface states, TI is supposed to open up innovative directions for future technological applications in spintronics and quantum computing as well. [4,5]

Since the first discovery of a two dimensional TI in an HgTe based quantum well,[6,7] several other families of materials for three-dimensional (3D) TI have been proposed theoretically and recently studied experimentally.[8-16] For example, tetradymite semiconcuctors, such as $Bi_2Te_3$, $Bi_2Se_3$, and $Sb_2Te_3$ are confirmed to be 3D TI with a single Dirac-cone on the surface, where the bulk band gap is as large as 0.3 eV, making the room temperature application possible.[8-12] However, a clear shortcoming of $Bi_2Te_3$ family is that these materials cannot be made with coexisting magnetism, a much desired property for spintronic applications. Although doping can be used to achieve the magnetically ordered behavior,[17] this creates extra complexity in material growth and could introduce detrimental effects upon doping.

Recently, a new family of ternary half-Heusler compounds with 18 valence electrons has been predicted to be 3D TI with proper stain engineering.[18-21] Most importantly, it was proposed that in the half-Heusler family the topological insulator allows the incorporation of new properties such as superconductivity or magnetism. By assuming a constant volume of the unit cell, Xiao *et al.*[20] have calculated the band structure of half-Heusler compound LaPtBi under uniaxial strain. They find that LaPtBi under proper uniaxial strain is a topological insulating phase. In order to simulate at best the experimental condition for thin film growth, in the present work, we have made use of an accurate full-potential density-functional method to study systematically the band structures for half-Heusler compound LaPtBi with the $C1_b$ structure under tetragonal distortions without the assumption of a constant volume. We will show that LaPtBi in the compressive in-plane strain state is an excellent topological insulator with a very large band gap of 0.3 eV. In addition, they will be proved to be mechanically stable and approximately 25-30 meV per formula unit higher in energy than the corresponding ground-state phases and, therefore, would be grown epitaxially in the form of thin films for spintronic applications.

The band-structure calculations in this work were performed using full-potential linearized augmented plane-wave method,[22] implemented in the package WIEN2K.[23] The exchange correlation of electrons was treated within the local spin density approximation (LSDA) including Spin-orbital coupling (SOC). Meanwhile, a 17×17×17 *k*-point grid was used in the calculations, equivalent to 5000 k points in the first Brillouin zone. Moreover, the muffin-tin radii of the atoms are 2.5 a.u, which are generated by the system automatically.

In order to determine the equilibrium lattice constant and find how the total energy varies with the lattice constant, we have performed structural optimizations on LaPtBi with $C1_b$ structure. In Figure 1, we show the total energy as a function of lattice constant. The equilibrium lattice constant $a_{eq}$ was determined by total energy minimization. It is found that LaPtBi with the lattice constant of 6.90Å shows the lowest energy, which is a little larger than that of the experimentally reported lattice parameter of 6.83 Å for bulk LaPtBi.[24] The inset of Fig. 1 shows the fully relativistic energy band structure of LaPtBi, which features a distinctive inverted band order at the Γ point, where the *s*-like $\Gamma_6$ states lie below the fourfold degenerate *p*-like $\Gamma_8$ states. Away from the Γ point, the valence band and conduction bands are well separated without crossing each other. Since the band inversion occurs only once throughout the Brillouin zone and therefore, LaPtBi is a topologically nontrivial phase in its unstrained state. Our result is consistent with previous calculations of ref. 20. Moreover, we should point out that the fourfold degeneracy of the $\Gamma_8$ states is protected by the cubic symmetry of the $C1_b$ structure.

For the investigation of the effects of tetragonal distortion on the topological band structures of LaPtBi, we next carried out a series of calculations of the compressive (tensile) strained LaPtBi using an reduced (increased) theoretical in-plane lattice constant *a* of a calculation unit cell with respect to the theoretical equilibrium lattice constant of LaPtBi, by leaving the *c*-axis unconstrained (free to relax). Here, in-plane strain was simulated by tetragonal lattice distortions using $a = \lambda \cdot a_{eq}$ with λ=0.94, 0.96,…,1.04, 1.06. The equilibrium value *c* was then determined by total energy minimization with fixed *a*. The total energies as a function of equilibrium *c* obtained from the calculations are shown in Figure 2(a). It is immediately clear that the cubic (λ=1.00) situation is the most stable state with minimum energy. Moreover, as shown in Fig. 2(b), the total energy of relaxed-volume calculations for the case of in-pane strain of ±2% is only increased by 25-30 meV, i.e., about the typical thermal energy at room temperature. Compared to the energies involved with homogenous lattice compression/expansion, such an energy scale is rather small. In the case of a Heusler thin film growing on a single crystalline substrate, in fact, the structure of film has to be considered as very soft with respect to small non-relaxed tetragonal distortions. According to our results, a change in the lattice parameters of LaPtBi thin film with varying growth conditions or annealing temperature can be caused by epitaxial matching with the substrates.

For comparison, In Fig. 2(b), we also show the total energy as a function of the distortion parameter $\lambda$ for the constant-volume (filled symbols) calculations. We find that the total energy of the fully relaxed-volume calculations is lower than that of the constant-volume one, which in turn strongly indicates that the assumption of a relaxed-volume is more reasonable. Moreover, as shown in Fig. 2 (c), the relaxed-volume is not constant upon distortion and increases monotonic with the increasing of $\lambda$. The change rate of volume is only about 1.25%, which means the contribution to the volume change would be more prominent from ab plane than c.

Further insight into the influence of tetragonal distortion on the band structures of LaPtBi is shown in Fig. 3. For clarity, we only give out the cases of $\lambda$ from 0.96 to 1.04. As expected, the in-plane strain lifts the degeneracy of the $\Gamma_8$ states and opens up an energy gap around $\Gamma$ point, while leaving the inverted band order intact. The most striking property here is the different behavior if the lattice is compressed ($\lambda$<1) or expanded ($\lambda$>1). With increasing compression, a global band gap is opened by the in-plane strain when the lattice is compressed with the Fermi energy lies in between ($\lambda$=0.96). However, in the expansion case, although a global band gap is opened, the Fermi energy now shifts to the valance bands, making the material effectively a topologically trivial phase ($\lambda$=1.04). Our results for the strained LaPtBi clearly suggest that epitaxial strain encountered during epitaxial growth of films can result in electronic topological transition from zero-gap semiconductors or topologically trivial phases to topological insulator states.

From the unstrained cubic structure to the in-plane strained tetragonal structure, the original $\Gamma_8$ states split into states with $\Gamma_7$ and $\Gamma_6$ symmetry, which typically form the top set of the valence bands and the bottom set of the conduction bands. We therefore define the strain-induced bulk band strength $\Delta E$ as the energy difference between the conduction band minimum and the valence band maximum at the $\Gamma$ point. In Fig. 4 (a) and (b), we show $\Delta E$ and the inversion strength, define as energy difference between the valence band maximum and the *s*-orbital originated $\Gamma_6$ at the $\Gamma$ point, as a function of the distortion parameter $\lambda$. We find that the compressive in-pane strain ($\lambda$<1) seems to have higher efficiency on opening the bulk band than that of tensile one ($\lambda$>1), while the inversion strength increases with the increasing $\lambda$. For the case of LaPtBi with relaxed-volume ($\lambda$=0.94), a strain-induced bulk band gap of 0.3 eV can be realized in LaPtBi with relaxed-volume. This is significantly larger than typical band gaps opened by the uniaxial strain in LaPtBi with a constant volume, [20] and is also well above the energy scale of room temperature. Thus, when dealing with gap-engineering technique by applying strain in half-Heusler compounds, it is necessary to consider relaxed geometries.

In the present study, we have shown by *ab-initio* calculations that the cubic crystal structure of LaPtBi is easily for tetragonal distortion, which we therefore expect to occur in epitaxially grown thin films. In comparison with calculations that assume a constant-volume, we find that the assumption of a relaxed-volume is more reasonable due to the energetically favorable. It is shown, that the bulk band gap as well as the Fermi level can be tuned by small lattice distortions in LaPtBi. We hope that the results presented here will motivate further experimental investigation of the electronic topological transition in LaPtBi or other half-Heusler compounds.

This work was supported by National Natural Science Foundation of China (Grant Nos. 51071172, 51021061, and 51025103).

## Figure captions:

FIG. 1. (Color on line) Total energy as a function of the relaxation of the cubic lattice constant *a*. For the minimum determination more points were calculated, which are not shown here. The inset shows the band structures of LaPtBi. The lattice constants are taken from the optimized values, i.e., at the total energy of the relaxed cubic configuration with $a_{eq}$=6.90Å.

FIG. 2. (Color on line) (a) Total energy as a function of the tetragonal lattice constant c, indicated for the different distortion parameters $\lambda$ ($\lambda = a/a_{eq}$). (b) Total energy for relaxed (opened circles) and constant (filled circles) unit cell volume calculations. (c) Relaxed unit cell volume as a function of the distortion parameter $\lambda$.

FIG. 3. (Color on line) Band structures of LaPtBi in different $\lambda$-values with relaxed c; The inset in (a) shows the energy gap induced by tetragonal distortion and $\Delta$E indicates the opened band strength.

FIG. 4. (Color on line) (a) The opened band strength $\Delta$E and (b) topological band inversion strength as a function of the distortion parameter $\lambda$. Here, the inversion strength is defined as the energy difference between the valence band maximum and the *s*-orbital originated $\Gamma_6$ at the $\Gamma$ point.

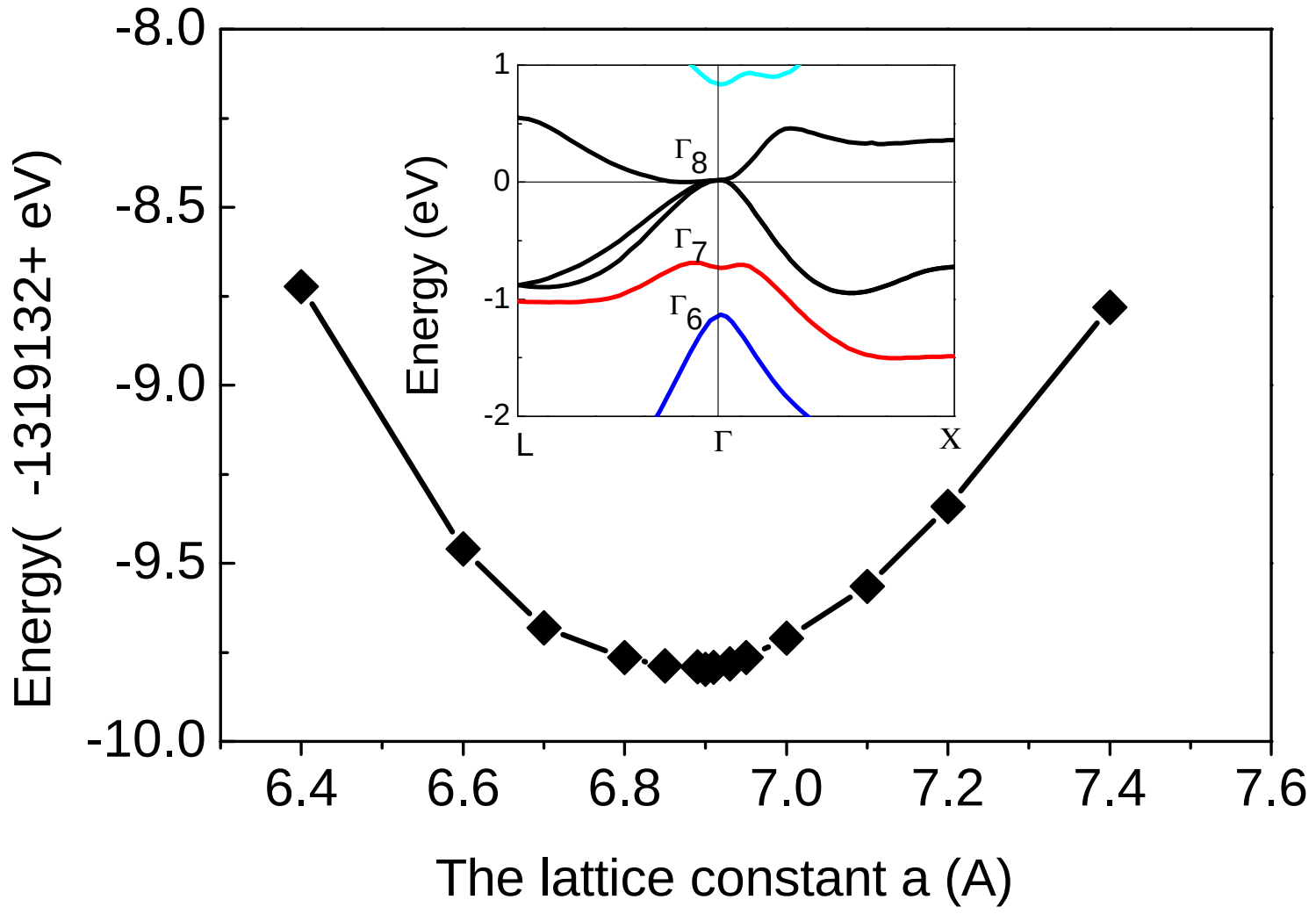

FIG. 1.

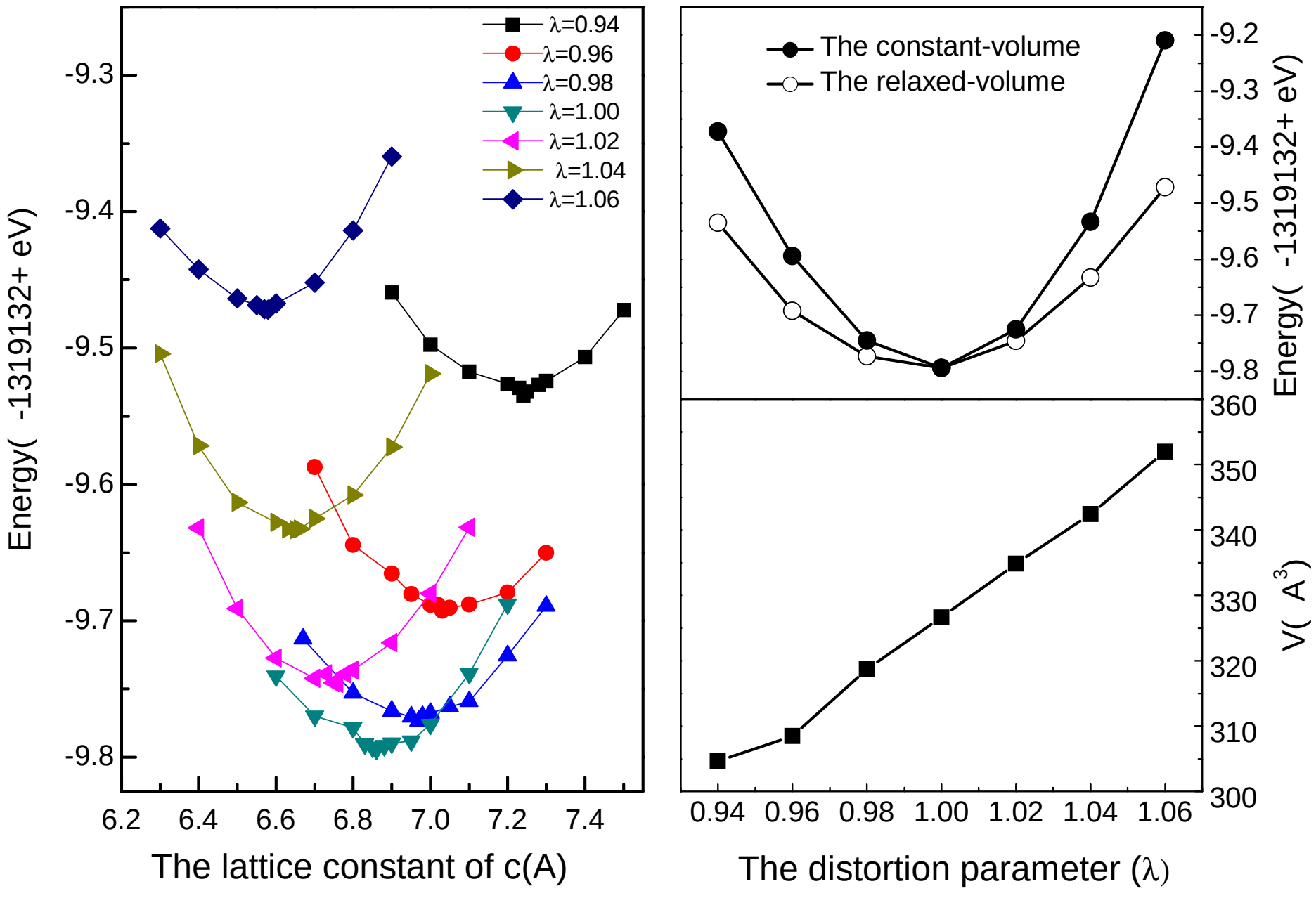

FIG. 2

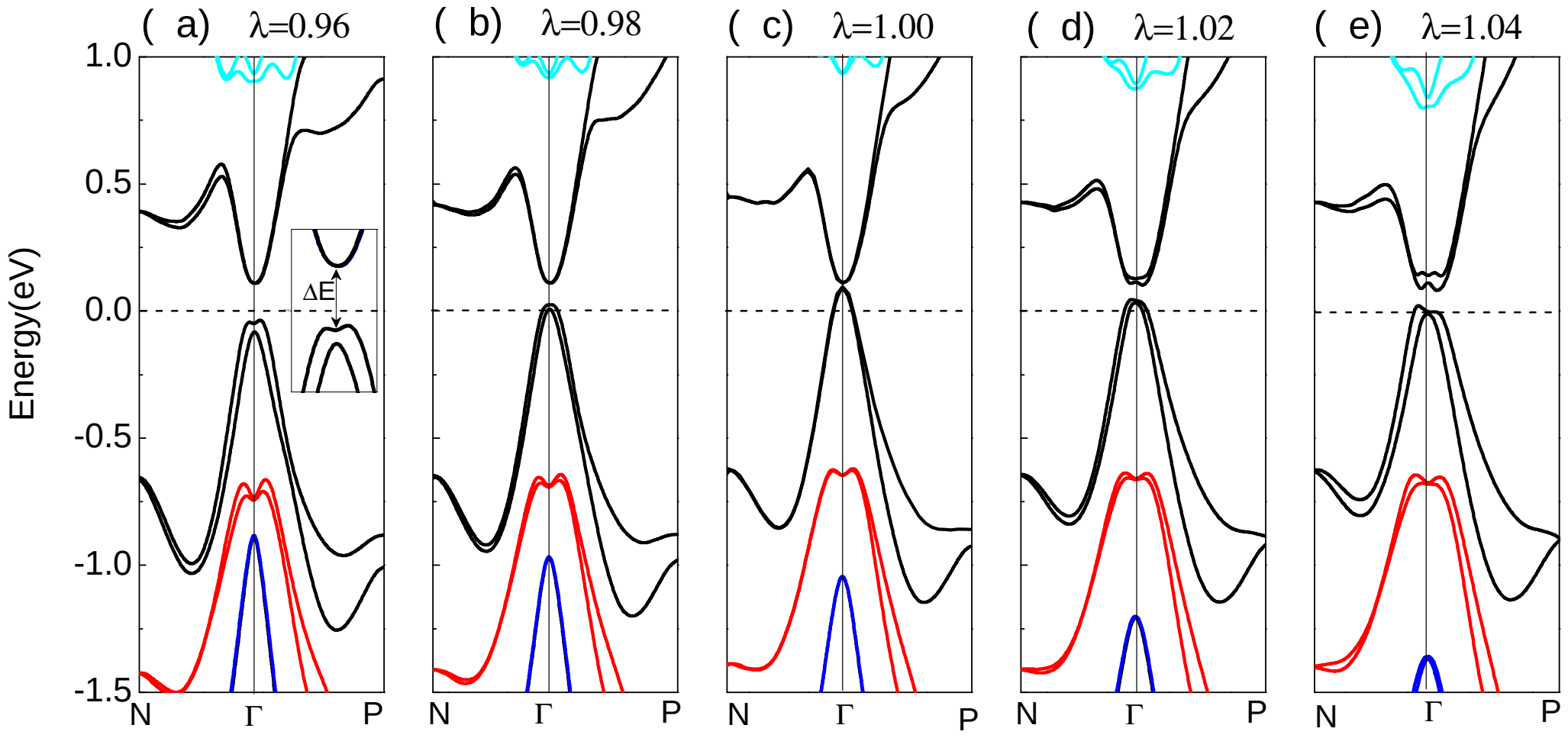

FIG. 3

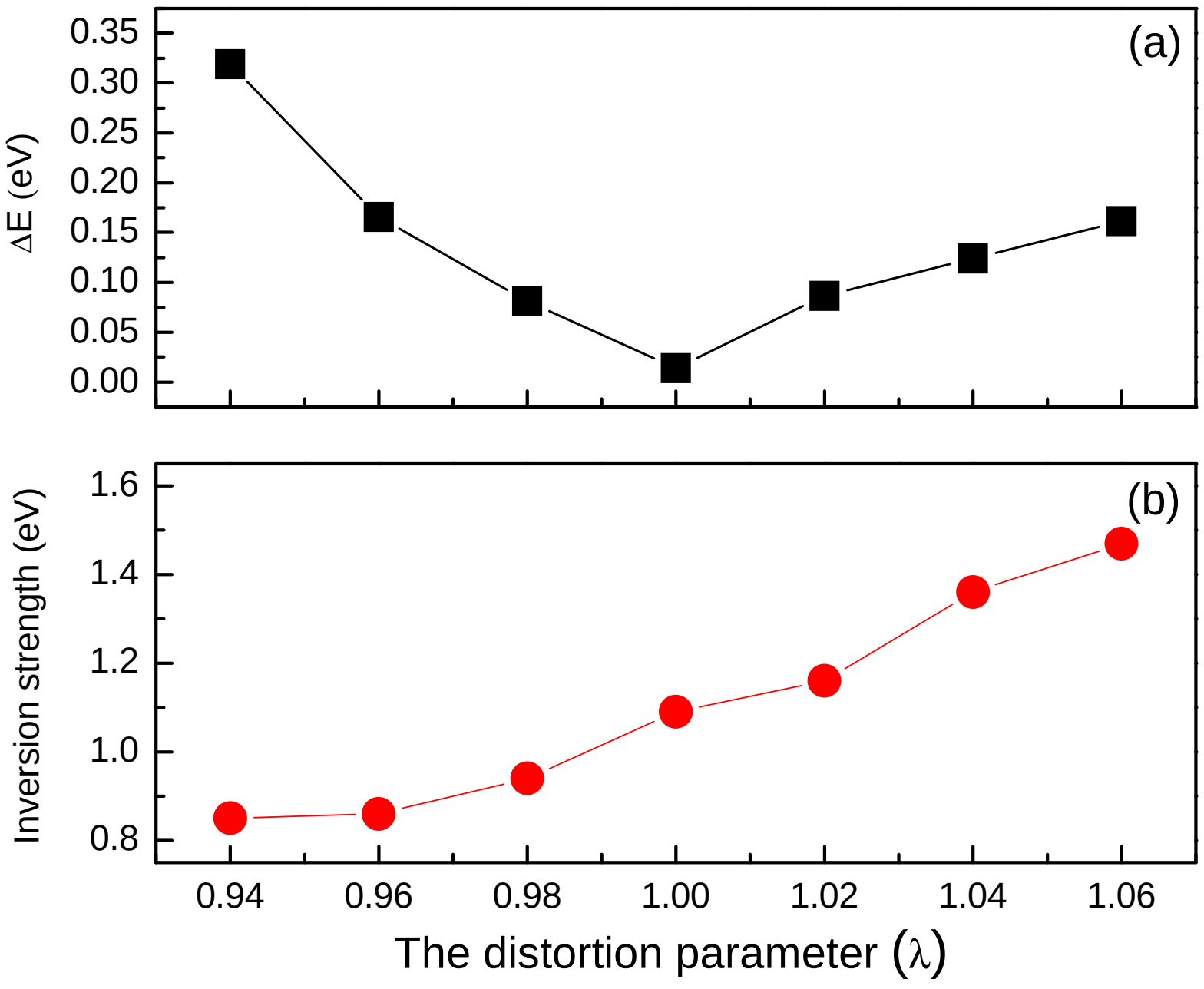

FIG. 4